\documentclass{llncs}

\usepackage[english]{babel}		
\usepackage[pdftex]{graphicx}	
\usepackage[hyphens]{url}
\usepackage{authblk}
\usepackage{color}
\usepackage{acro}
\usepackage{newclude}
\usepackage[utf8]{inputenc}
\usepackage[acronym,toc]{glossaries}
\usepackage[normalem]{ulem}  
\usepackage{subfig}
\usepackage{tikz}
\usepackage{tkz-tab}
\usepackage{pgfbaseimage}
\usetikzlibrary{shapes,snakes,shadows,calc,automata,positioning, backgrounds}
\usepackage{adjustbox}

\tikzstyle{abstract}=[rectangle, draw=black, rounded corners, fill=blue!40, drop shadow,
        text centered, text=white, text width=2.5cm]
\tikzstyle{abstract2}=[rectangle, draw=black, rounded corners, fill=blue!40, drop shadow,
        text centered, text=white, text width=2.5cm, minimum height = 2.5cm]
\tikzstyle{policy}=[rectangle, draw=black, rounded corners, fill=red!10, text centered, text=red, text width=1.5cm]
\tikzstyle{network}=[rectangle, draw=black, rounded corners, fill=green!40!black!60, drop shadow, text centered, text=white, text width=1.5cm, minimum height = 0.8cm]
\tikzstyle{MNO}=[rectangle, draw=black, rounded corners, fill=blue!40!black!60, drop shadow, text centered, text=white, text width=1.5cm, minimum height = 0.8cm]

\makeatletter
\newcommand\footnoteref[1]{\protected@xdef\@thefnmark{\ref{#1}}\@footnotemark}
\makeatother

\title{An Overview of GSMA's M2M Remote Provisioning Specification}
\author{Maxime Meyer\inst{1} \and Elizabeth A. Quaglia\inst{2} \and Ben Smyth\inst{3}}
\institute{Innovation Lab, Vade Secure, Canada
\and Information Security Group - Royal Holloway, University of London, UK
\and Interdisciplinary Centre for Security, Reliability and Trust,\authorcr University of Luxembourg, Luxembourg
}

\date{\today}

\DeclareAcronym{eid}{
  short = EID,
  long = eUICC-ID,
  class = abbrev,
  sort = eid
}

\DeclareAcronym{sm}{
  short = SM,
  long  = Subscription Manager,
  class = abbrev,
  sort = sm
}

\DeclareAcronym{sr}{
  short = SM-SR,
  long  = Subscription Manager - Secure Routing,
  class = abbrev,
  sort = smsr
}
\DeclareAcronym{ecasd}{
  short = ECASD,
  long  = eUICC Controlling Authority Security Domain,
  class = abbrev,
  sort = ecasd
}
\DeclareAcronym{dp}{
  short = SM-DP,
  long  = Subscription Manager - Data Preparation,
  class = abbrev,
  sort = smdp
}
\DeclareAcronym{sim}{
  short = SIM ,
  long  = Subscriber Identity Module ,
  class = abbrev,
  sort = sim
}
\DeclareAcronym{usim}{
  short = USIM ,
  long  = Universal Subscriber Identity Module ,
  class = abbrev,
  sort = usim
}
\DeclareAcronym{apdu}{
  short = APDU,
  long  = Application Protocol Data Unit,
  class = abbrev
}
\DeclareAcronym{m2m}{
  short = M2M,
  long  = Machine to Machine ,
  class = abbrev
}
\DeclareAcronym{gsm}{
  short = GSM,
  long  = Global System for Mobile Communication,
  class = abbrev
}
\DeclareAcronym{umts}{
  short = UMTS,
  long  = Universal Mobile Telecommunications System,
  class = abbrev
}
\DeclareAcronym{lte}{
  short = LTE,
  long  = Long-Term Evolution,
  class = abbrev
}
\DeclareAcronym{gsma}{
  short = GSMA,
  long  = Groupe Sp\'eciale Mobile Association,
  class = abbrev
}
\DeclareAcronym{etsi}{
  short = ETSI,
  long  = European Telecommunications Standards Institute,
  class = abbrev
}
\DeclareAcronym{euicc}{
  short = eUICC,
  long  = embedded UICC,
  class = abbrev,
  sort = euicc
}
\DeclareAcronym{uicc}{
  short = UICC,
  long  = Universal Integrated Circuit Card,
  class = abbrev
}
\DeclareAcronym{icc}{
  short = ICC,
  long  = Integrated Circuit Card,
  class = abbrev
}
\DeclareAcronym{mno}{
  short = MNO,
  long  = Mobile Network Operator,
  class = abbrev
}
\DeclareAcronym{iot}{
  short = IoT,
  long  = Internet of Things,
  class = abbrev
}
\DeclareAcronym{se}{
  short = SE,
  long  = Secure Element,
  class = abbrev
}
\DeclareAcronym{tee}{
  short = TEE,
  long  = Trusted Execution Environment,
  class = abbrev
}
\DeclareAcronym{ree}{
  short = REE,
  long  = Rich Execution Environment,
  class = abbrev
}
\DeclareAcronym{tpmm}{
  short = TPM Mobile,
  long  = Trusted Platform Module Mobile,
  class = abbrev
}
\DeclareAcronym{tpm}{
  short = TPM,
  long = Trusted Platform Module,
  class = abbrev
}
\DeclareAcronym{tcg}{
  short = TCG,
  long = Trusted Computing Group,
  class = abbrev
}
\DeclareAcronym{tc}{
  short = TC,
  long = Trusted Computing,
  class = abbrev
}
\DeclareAcronym{cots}{
  short = COTS,
  long = Commercial Of The Shelf,
  class = abbrev
}

\DeclareAcronym{imsi}{
  short = IMSI,
  long = International Mobile Subscriber Identity,
  class = abbrev
}
\DeclareAcronym{eum}{
  short = EUM,
  long = eUICC Manufacturer,
  class = abbrev,
}
\DeclareAcronym{prof}{
  short = Profile,
  long = Profile,
}
\DeclareAcronym{isdr}{
  short = ISD-R,
  long = Issuer Security Domain Root,
  class = abbrev,
}
\DeclareAcronym{isdp}{
  short = ISD-P,
  long = Issuer Security Domain Profile,
  class = abbrev,
}
\DeclareAcronym{ci}{
  short = CI,
  long = Certificate Issuer,
  class = abbrev,
}
\DeclareAcronym{eis}{
  short = EIS,
  long = eUICC Information Set,
  class = abbrev,
}

\newtoggle{ArxivPaper}
\togglefalse{ArxivPaper}
\iftoggle{ArxivPaper}{
}

\begin{document}
\maketitle

\begin{abstract}

M2M devices are ubiquitous, and there is a growing tendency to connect such devices to mobile networks. 
Network operators are investigating new solutions to lower their costs and 
to address usability issues.
Embedded SIM cards with remote provisioning capability are one of the most promising solutions.
GSMA, the leading consortium on mobile network standards, has proposed a specification for such an embedded SIM card, called eUICC. 
The specification describes eUICC architecture 
and a remote provisioning mechanism.
Embodiments of this specification have the potential to disrupt 
the telecommunications market: eUICCs will be shipped to device manufacturers and then remotely provisioned with a subscription, 
whereas (currently) SIMs must be provisioned prior to shipping.
In this article, we present a comprehensive overview of GSMA's specification and its motivation. In particular, we describe the technology and the protocols involved in remote provisioning. 

\end{abstract}

\setcounter{footnote}{0}
\section{Introduction}

\ac{m2m} devices communicate without human intervention.
They are ubiquitous and have diverse applications, ranging from smart (e.g.,  smart-watches, -meters, and -cars) and medical devices, to ATMs and vending machines, for instance. 
A growing number of devices communicate over cellular networks~\cite{GMSAint}, 
 relying on infrastructures owned by \acl{mno}s (\acs{mno}s), 
who restrict network access to subscribers.

To receive service, \ac{m2m} devices authenticate to MNOs,
using an authenticated key-exchange protocol. For 3G and 4G networks, the Authentication and Key Agreement protocol (AKA) \cite{ERICSSON} is used,%
which allows a secure, over-the-air channel to be established
between the device and the network.
This is achieved using several cryptographic algorithms, a unique identifier (for each subscriber), and a symmetric key.%
\iftoggle{ArxivPaper}{
	\footnote{\label{note1}The implementation of AKA's algorithms is left to MNOs, but, for 3G, 3GPP proposed an implementation of those algorithms based on AES, called MILENAGE \cite{milenage} and ETSI proposed an alternative implementation, based on the Keccak sponge function, called TUAK \cite{tuak}. The AKA protocol security guaranties have been analyzed \cite{alt2016cryptographic}.}
}\

\label{subsec:mmf2distributionchannel}

\begin{figure}[!h]
	\centering     
  	\resizebox{\linewidth}{!}{
			\begin{tikzpicture}[framed]
	
	  			\node (EUM) at (0,0) {};
  				\node (MNO) at (3,0) {};
	  					
  				\node (M) at (6,0) {\includegraphics[width=50pt]{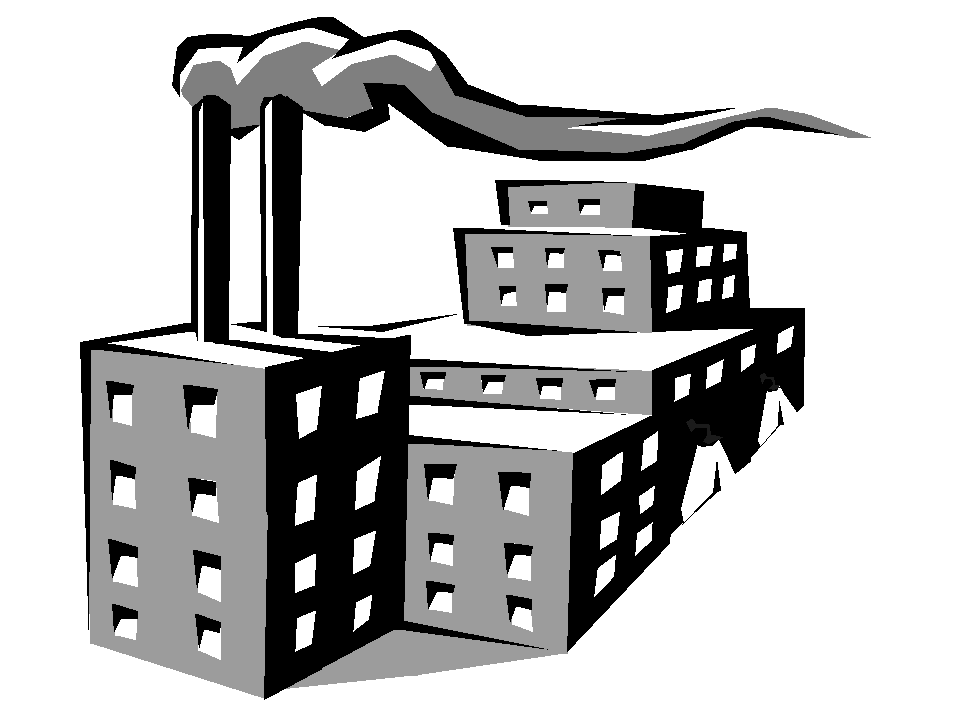}};
  				\draw ($ (M) + (0,-0.7) $) node[below] {M2M Manufacturer};	
				\node (M2M) at (9,0) {\includegraphics[width=50pt]{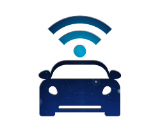}};
  				\draw ($ (M2M) + (0,-0.7) $) node[below] {M2M Device};

				\draw (MNO) node[MNO] {MNO};
				\draw[black!60!green, very thick] ($ (MNO) + (1,1) $) node {2};
			
				\draw[->, black!60!green, very thick] ($ (MNO) + (0.75,0.2) $) arc (-80:200:3mm) ;
				\draw[-stealth, black!60!green, very thick] ($ (MNO) + (0.75,0) $) -- node[above] {3} (M);	
				\draw (MNO) node[MNO] {MNO};
				\draw[-stealth, black!60!green, very thick] ($ (EUM) + (1.25,0) $) -- node[above] {1} ($ (MNO) + (-0.75,0) $);	
				\draw[-stealth, black!60!green, very thick] (M) -- node[above] {4} (M2M);
		  		\draw (EUM) node[MNO, text width=2cm, minimum width = 2.7cm] {MFF2\ \ SIM \\ Manufacturer};
	  		\end{tikzpicture}
	}
	\caption{MFF2 SIM linear distribution \& subscription model.} \label{fig:linearLifeCycle}
\end{figure}
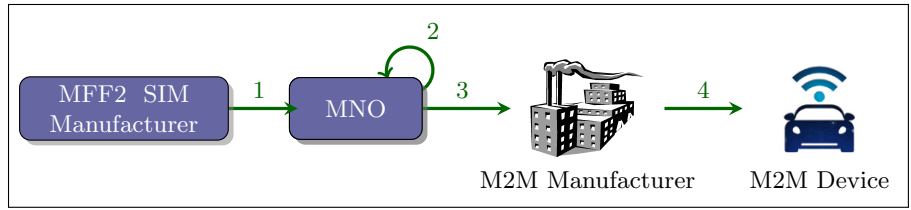

Historically, M2M devices have used dedicated smart cards, called \acl{sim}s (\acs{sim}s), to store the secrets needed to authenticate to and access operators' networks.
The SIMs used by M2M devices to access cellular networks are specified by the European Telecommunication Standards Institute (ETSI). 
They are known as Machine Form Factor (MFF2) \acs{sim}s \cite{ETSIMFF} and are distributed using the following linear chain (illustrated in Figure \ref{fig:linearLifeCycle}):

\begin{enumerate}\label{enum:test}
	\item MFF2 SIM cards are fabricated by manufacturers and shipped to MNOs.
	\item MNOs personalize MFF2 SIMs by installing \textit{subscription data}
\iftoggle{ArxivPaper}{
\footnote{Subscription data refers to the set of data and applications necessary to allow a device to connect and use a network. This includes, but is not limited to, implementations of algorithms required by AKA, a symmetric key, a network access application compatible with 3G or 4G networks such as USIM application and a file system and the parameters needed to provide contracted services to the device.} 
}
{
} over a physical connection.
\iftoggle{ArxivPaper}{}{(Subscription data refers to data and applications necessary to allow a device to connect and use a network, including a symmetric key, network parameters, a network access application (NAA), and
an implementation of AKA.)\footnote{Alternatively, personalization is often done by the MFF2 manufacturer who ships SIM cards to MNOs along with the SIMs personalization information.}}

	\item Personalized MFF2 SIMs are issued to M2M device manufacturers.
	\item Device manufacturers embed MFF2 SIMs inside devices. (In Figure \ref{fig:linearLifeCycle}, the wireless logo appears  on top of the M2M device to illustrate that the device has a subscription.)

\end{enumerate}

\noindent
Devices are then shipped to the relevant distributors, wholesalers, and retailers.

\tikzstyle{channel}=[rectangle, draw=black, rounded corners, fill=blue!40, drop shadow, text centered, text=white, text width=1cm, minimum height = 0.5cm]
     
MFF2 SIMs suffer shortcomings including the following:

\begin{itemize}

\item{Subscription data must be installed using a physical channel,}
which increases costs and lead times.
\item{MFF2 SIMs can only support a single subscription from a single MNO.}
Hence, a product intervention is required to change a subscription, which is
costly for MNOs \cite[p.13]{GSMAINT} and manufacturers, 
especially for devices in 
remote, inaccessible locations, e.g., underwater sensors used by offshore oil platforms.

\item{Changing MFF2 SIMs is often a difficult operation, because to do so devices must typically be unsealed.}
Indeed, most remote M2M devices are sealed, either to prevent theft, or to protect its components when the device is operating in harsh environmental conditions.

\end{itemize}

\noindent
Stakeholders are seeking new technologies to  solve these shortcomings.

\iftoggle{ArxivPaper}{%
Several consortia, including ETSI and GSMA,\footnote{Groupe Sp\'eciale Mobile Association (GSMA) is a consortium of companies forming a membership association created to promote and represent the interest of MNOS worldwide. Currently more than 800 operators are full members of GSMA.}
\textcolor{blue}{Is it worth mentioning that consortia other than ETSI and GSMA exist? If not, then just write 
\emph{consortia ETSI and GSMA}. (You can just revise the if-then-else.)}
}{%
ETSI and GSMA%
}
are investigating \emph{remotely re-programmable} \ac{sim}s to address 
shortcomings.
The first specification for a remotely re-programmable \ac{sim} was proposed by ETSI in working document TS 103.383~\cite{ETSI}.
This specification introduces an embedded \acl{uicc} (eUICC) and the core requirements to enable remote provisioning.
\acs{gsma} have extended that specification~\cite{GSMA1}, and 
have also released supporting documents~\cite{GSMABPR}
, providing use cases and test cases  for such an eUICC.
GSMA's specification has ``received strong industry support" from several key operators, including AT\&T, Deutsche Telekom, Orange, and Vodafone.

\paragraph{Contribution and motivation.}

This article introduces and explains GSMA's ``Remote Provisioning Architecture for Embedded UICC" specification~\cite{GSMA1}, with a focus on the smart card architecture, and on the remote provisioning mechanisms.
Our work aims at facilitating the understanding of GSMA's specification by presenting 
a comprehensive overview of GSMA's specification, and 
 of their M2M specification.
GSMA's specification for \acs{euicc}s is marketed and considered by many operators and device manufacturers as the leading proposition for the next SIM technology 
 that makes an attempt to meet the requirements for both remote provisioning and re-programmability of the SIM.
As such, we believe it is of paramount importance to explain GSMA's specification and the motivations behind it.
As part of our contribution, we present detailed visuals of eUICCs and their environment.

\paragraph{Structure.}
We illustrate the changes to the SIM distribution and subscription model
(Section 2), describe the newly proposed ecosystem for M2M devices (Section~3), and present the eUICC architecture (Section 4). Moreover,  
we explain the core functions and mechanisms behind remote provisioning
(Section 5), and close with a brief conclusion (Section~6). (For 
reference, a list of abbreviations appears towards the end of this 
article.)

\section{eUICC distribution chain and subscription model}
\label{sec:model}

\begin{figure}
	\centering	
  	\resizebox{\linewidth}{!}{
  		\subfloat[Distribution chain]{%
			\begin{tikzpicture}[framed]

		  		\node (EUM) at (0,0) {};
				\node (SM) at (3,0) {};	  				
	  			
	  			\node (M) at (0, -3) {\includegraphics[width=50pt]{manufacture.png}};
				\draw ($ (M) + (0,-0.7) $) node[below] {Device Manufacturer};	
				\node (M2M) at (3,-3) {\includegraphics[width=40pt]{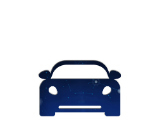}};
				\draw ($ (M2M) + (0,-0.7) $) node[below] {Device};

				\draw (SM) node[network] {SM};
				
				\draw[black!60!green, very thick] ($ (EUM) + (0.75,0.8) $) node {1};
				\draw[->, black!60!green, very thick] ($ (EUM) + (0.5,0.4) $) arc (-10:190:4mm) ;
				
				\draw[-stealth, black!60!green, very thick] ($ (EUM) + (0.75,0) $) -- node[above] {2} ($ (SM) + (-0.75,0) $);									
				\draw[-stealth, black!60!green, very thick] ($ (EUM) + (0,-0.3) $) -- node[right] {3} (M);
				\draw[-stealth, black!60!green, very thick] (M) -- node[above] {4} (M2M);	
	            \draw (EUM) node[MNO] {EUM};
	
			\end{tikzpicture}
		}
  		\subfloat[Subscription model]{%
			\begin{tikzpicture}[framed]

	  			\node (MNO) at (5,-1) {};
				\node (SM) at (3,0) {};	  				
	  			\draw[white, very thick] ($ (SM) + (0.75,0.8) $) node {1};

				\node (M2M) at (3,-3) {\includegraphics[width=40pt]{car}};
				\draw ($ (M2M) + (0,-0.7) $) node[below] {Device};	
				
				\draw[-stealth, black!60!green, very thick] ($ (SM) + (0,-0.3) $)-- node[right] {6} (M2M);
				\draw (SM) node[network] {SM};
				\draw[black!60!green, very thick] ($ (MNO) + (-0.8,0.8) $) node {5};
				\draw[-stealth, black!60!green, very thick] ($ (MNO) + (-0.75,0.3) $) -- ($ (SM) + (0.75,-0.3) $);				
				\draw (MNO) node[MNO] {MNO};
				
				\draw[white, very thick] ($ (SM) + (0.75,0.75) $) node {1};

			\end{tikzpicture}
		}
		}
	\caption{eUICC distribution chain \& subscription model}\label{fig:euiccChannel}		
\end{figure}

%
Remote provisioning allows consumers to buy M2M devices before subscribing to an MNO.
This leads to an evolution of the linear distribution model (used by SIMs), since distribution and subscription can be separated.
To instantiate remote provisioning, GSMA introduces a \ac{sm}, which acts as an intermediary between MNOs and eUICCs.
This removes MNOs from distribution and results in a new chain, described as follows (and which we illustrate in Figure \ref{fig:euiccChannel} (a)):
\begin{enumerate}
	\item eUICCs are fabricated by \acl{eum}s (\acs{eum}s).
	\item EUMs create an \textit{eUICC Information Set (EIS)} file for each eUICC describing their characteristics,\footnote{These are permanent characteristics, such as the EUM identity or the eUICC production date signed by the EUM.} and send the file to the SM.
	\item eUICCs are shipped to M2M device manufacturers.
	\item M2M device manufacturers embed eUICCs inside devices.
\end{enumerate}


\noindent
As before, devices are shipped to the relevant distributors, wholesalers, and retailers.

The eUICC distribution chain reduces production costs for EUMs and for M2M manufacturers.
Indeed, as MNOs are removed from the distribution chain, eUICCs are manufactured without subscription and can be shipped directly to device manufacturers who can embed the eUICC.
Furthermore, eUICCs can be embedded deep inside devices as they will be remotely managed.
MNO costs are also reduced, as they need not ship SIMs to device manufacturers. Finally, eUICCs need not be personalized over a physical channel. 

The new distribution chain leads to a new subscription model, whereby subscriptions are purchased as follows (and which we illustrate in Figure \ref{fig:euiccChannel} (b)):
\begin{enumerate}
\item[5.] The owner of a M2M device subscribes to an MNO and the MNO sends a request to download the related subscription to the SM.
\item[6.] The SM writes subscription data corresponding to the MNO to the eUICC application, called a \emph{Profile}, 
and remotely installs this profile on the eUICC present in the M2M device.
\end{enumerate}


\noindent
It follows that devices and subscriptions can be purchased separately, benefiting buyers who can independently leverage the device and subscription markets. 

We next describe eUICCs and the mechanisms behind remote provisioning as well as the associated protocols.
\section{eUICC architecture}
\label{sec:euicc}


MFF2 SIMs are compatible with GlobalPlatform's\footnote{GlobalPlatform is a consortium maintaining multi-application smart card standards.} smart card specification standard  \cite{GPC}.\footnote{Markantonakis and Mayes present an overview of GlobalPlatform's smart card specification \cite{GPCoverview}
}
This standard specifies a general architecture for multi-application platforms, application lifecycle, platform management, and input-output channels between the card and external entities.
In order to ease transition from MFF2 SIMs to eUICCs and for backward compatibility, eUICCs are also compliant with GlobalPlatform's standard.

It follows that eUICCs build upon the architecture of GlobalPlatform-compliant smart cards\iftoggle{ArxivPaper}{ (represented in Figure \ref{fig:GPsmartcard}).}{.}
As such, eUICCs contain a runtime environment, a framework (called OPEN) 
that interacts with different applications and \textit{Security Domains}, i.e.,  privileged applications that act as on-card representatives of off-card authorities and that can establish secure channels and handle some management functions.
Each application, including security domains, depends on a parent (or master) application.
This association permits parent-child communication, while other communication requires authorization from the parent.
Messages received and processed by the eUICC respect the \ac{apdu} format specified by GlobalPlatform standard.

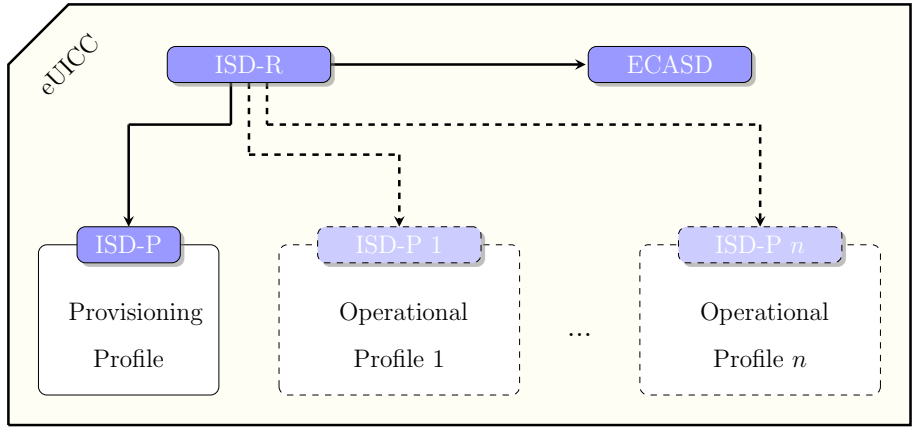
\begin{figure}
    \centering
	\resizebox{\linewidth}{!}{
    	\begin{tikzpicture}[font=\fontsize{12}{22.4}\selectfont]

			\draw[very thick,fill = yellow!5] (0,0) -- (0,6) -- (1,7) -- (15,7) -- (15,0) -- (0,0);
			\node[rotate=45] at (1,6) {eUICC};

			\draw[rounded corners, fill = white] (0.5,0.5) rectangle (3.5,3);
			\draw[rounded corners, dashed, fill = white] (4.5,0.5) rectangle (8.5,3);
			\draw[rounded corners, dashed, fill = white] (14.5,0.5) rectangle (10.5,3);
			
			\node (P1) at (2,3) {};
			\node (P0) at (6.5,3) {};
			\node (P2) at (12.5,3) {};
			\node (R) at (4, 6) {};
			\node (CASD) at (11,6) {};

			\node at (9.5, 1.5) {\Large{...}};
			\node[text width = 2cm, text centered] at (2, 1.5) {Provisioning Profile};
			\node[text width = 2cm, text centered] at (6.5, 1.5) {Operational Profile 1};
			\node[text width = 2cm, text centered] at (12.5, 1.5) {Operational Profile $n$};							

			\draw[-stealth, very thick] ($ (R) + (1.3,0) $) -- ($ (CASD) + (-1.4,0) $);
			\draw[very thick, dashed] ($ (R) + (0.3,-0.3) $)	-- ($ (R) + (0.3,-1) $);			
			\draw[very thick] ($ (R) + (-0.3,-0.3) $)	-- ($ (R) + (-0.3,-1) $);
			\draw[very thick, dashed] ($ (R) + (0,-0.3) $)	-- ($ (R) + (0,-1.5) $);
			\draw[very thick] ($ (R) + (-0.3,-1) $) -- (2,5);
			\draw[very thick, dashed] ($ (R) + (0,-1.5) $)	-- (6.5,4.5);
			\draw[very thick, dashed] ($ (R) + (0.3,-1) $)	-- (12.5,5);
			\draw[-stealth, very thick] (2,5) -- ($ (P1) + (0,0.3) $);
			\draw[-stealth, very thick, dashed] (12.5,5) -- ($ (P2) + (0,0.3) $);
			\draw[-stealth, very thick, dashed] (6.5,4.5) -- ($ (P0) + (0,0.3) $);		

			\draw (P1) node[abstract, minimum height = 0.6cm, text width = 1.5cm] {ISD-P};
			\draw (P0) node[rectangle, draw=black, rounded corners, fill=blue!20, drop shadow,
        text centered, text=white, text width=2.5cm, minimum height = 0.6cm, dashed] {ISD-P $1$};	
			\draw (P2) node[rectangle, draw=black, rounded corners, fill=blue!20, drop shadow,
        text centered, text=white, text width=2.5cm, minimum height = 0.6cm, dashed] {ISD-P $n$};		
			\draw (R) node[abstract, minimum height = 0.6cm] {ISD-R};
			\draw (CASD) node[abstract, minimum height = 0.6cm] {ECASD};
				
  		\end{tikzpicture}
  	}
	\caption{eUICC Architecture Overview. Solid boxes represent components installed during manufacture while dashed boxes represent components installed with remote provisioning. Lines represent the hierarchical association between components.} \label{fig:eUICCarchi}
\end{figure}

Compared to MFF2 SIMs that are tied to a single subscription, the eUICC architecture can support several operators simultaneously.
Each operator controls a profile on the eUICC. This profile is organised as an MFF2 SIM, i.e., the logical file system and internal application of an MFF2 SIM are preserved in the profile. The organisation of an eUICC, which we illustrate in Figure \ref{fig:eUICCarchi}, is now described.

As discussed in Section \ref{sec:model}, eUICCs are manufactured and shipped to the device manufacturer without subscriptions.
Some components and applications are installed at manufacture, while some are installed later on, relying on remote provisioning.
Freshly shipped eUICCs only contain the following elements (components represented with solid lines in Figure \ref{fig:eUICCarchi}):

\begin{itemize}
	\item An interface, the \emph{ISD-R}, for over-the-air communication with the subscription manager.
	
The \ac{isdr} provides an on-card communication interface accessible by the subscription manager.
The \ac{isdr} is the highest privileged security domain on the eUICC, and, as such, has management capabilities over all other eUICC applications, including security domains.
Remote commands sent to eUICCs are received and processed by the ISD-R, and then relayed to the targeted application, such as a profile.
The \ac{isdr} is equipped with a symmetric key \iftoggle{ArxivPaper}{($k_{80}$, see Section \ref{sec:channelproto}) }during eUICC's manufacture for establishing a secure communication channel with the subscription manager.

	\item A Controlling Authority, the \emph{ECASD}.

The \ac{ecasd} is responsible for authenticating remote parties using a public key infrastructure.
To achieve authentication, the \ac{ecasd} holds cryptographic data, including the public key of the certificate authority.
The \ac{ecasd} also holds the private key of the eUICC used to set up a secure channel (see Section \ref{sec:profile}).
The eUICC public key is stored in the \texttt{EIS} file stored by the subscription manager (see Section 2).

	\item A communication application to enable initial remote management of the eUICC.

The fabrication process of eUICCs includes granting them with minimal connectivity services to allow remote provisioning of MNO's profiles on the eUICC thereafter. 
For this purpose, a \emph{provisioning} profile (i.e., a profile containing an NAA set with minimal connectivity capabilities) is installed.
\end{itemize}

Once embedded inside a device, eUICCs are remotely provisioned with \emph{operational} profiles (i.e., a profile containing subscription data corresponding to an MNO).
eUICCs support many operational profiles throughout their lifetime.
Remote provisioning enables installation, management and deletion of those profiles.
Each profile is dependent on a dedicated security domain called an \ac{isdp}.
ISD-Ps are created by the ISD-R, which, despite management capabilities, is unable to modify their profile content, which should guarantee integrity of the operators' data stored in profiles.
By using dedicated ISD-Ps, profile data can only be accessed with the ISD-R's authorization, and is, as such, isolated from other profiles.

%
%
%
%
%
%
\section{Remote provisioning ecosystem}
\label{sec:remoteProvisioning}

In their specification, GSMA describe various communication channels between eUICCs and external entities (represented in Figure \ref{fig:remoteprovisioning}).
As communications between entities and the eUICC are wireless, GSMA has chosen to rely on certificates for authentication.
Thus, each entity involved in the remote provisioning framework is certified by an authority referred to as the \acs{ci} in GSMA's specification.

\tikzstyle{eUICC}=[rectangle, draw=black, rounded corners, fill=blue!40, drop shadow, text centered, text=white, text width=1.5cm, minimum height = 0.8cm]
\tikzstyle{network}=[rectangle, draw=black, rounded corners, fill=green!40!black!60, drop shadow, text centered, text=white, text width=1.5cm, minimum height = 0.8cm]
\tikzstyle{MNO}=[rectangle, draw=black, rounded corners, fill=blue!40!black!60, drop shadow, text centered, text=white, text width=1.5cm, minimum height = 0.8cm]
\tikzstyle{Profile}=[rectangle, draw=black, rounded corners, fill=blue!40!black!60, drop shadow, text centered, text=white, text width=1cm]

\begin{figure*}[!h]
	\centering     
  	\resizebox{0.8\linewidth}{!}{
  		\centering 
		\begin{tikzpicture}
		\tikzset{font={\fontsize{12pt}{12}\selectfont}}

			\node (MNO) at (0,0) {};
			\node (DP) at (2,1) {};
			\node (SR) at (5,1) {};
			\node (R) at (10,1) {};
			\node (E) at (13,2) {};
			\node (P) at (13,0) {};
			
			\draw[fill = yellow!5, dashed] (8.5,-0.6) -- (8.8,-1) -- (14.5,-1) -- (14.5,3) -- (8.5,3) -- cycle;
			\node[right, below] at (9.3,2.7) {\textit{eUICC}};

			\draw (MNO) node[MNO] {MNO};
			\draw (DP) node[network] {SM-DP};
			\draw (SR) node[network] {SM-SR};	
			\draw (R) node[eUICC] {ISD-R};	
			\draw (E) node[eUICC] {ECASD};	
			\draw (P) node[eUICC] {ISD-P};
			\node[Profile] (P1) at (13.2,-0.5) {$P_{MNO}$};

			\node at ($ (MNO) + (0.9,-0.4) $) {\includegraphics[width=18pt]{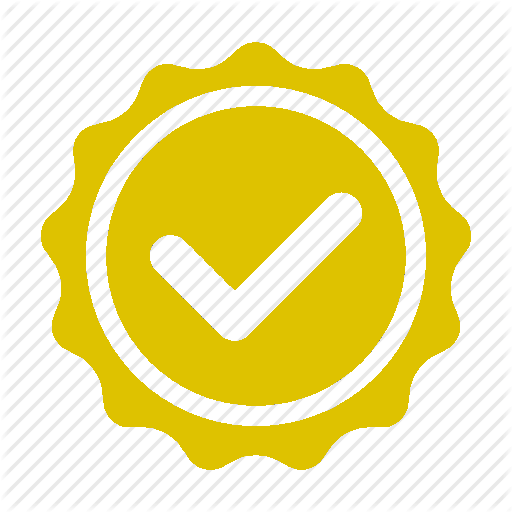}};
			\node at ($ (DP) + (0.9,-0.4) $) {\includegraphics[width=18pt]{certif.png}};
			\node at ($ (SR) + (0.9,-0.4) $) {\includegraphics[width=18pt]{certif.png}};
			\node at ($ (E) + (0.9,-0.4) $) {\includegraphics[width=18pt]{certif.png}};
			
			\draw[double, stealth-stealth, black!60!green, line width=0.7mm] ($ (SR) + (1,0) $) -- node[above] {1 (*)} ($ (R) + (-1,0) $);
			
			\draw[stealth-, dashed] ($ (R) + (1,0) $) -- (11.5,1);
			\draw[dashed] (11.5,2) -- (11.5,0);
			\draw[stealth-, dashed] ($ (P) + (-1,0) $) -- (11.5,0);
			\draw[stealth-, dashed] ($ (E) + (-1,0) $) -- (11.5,2);
			
			\draw[stealth-stealth, black!60!green] ($ (DP) + (1,0) $) -- node[above] {(*)} ($ (SR) + (-1,0) $);
			\draw[stealth-stealth, black!60!green] ($ (MNO) + (0,0.5) $) -- ($ (MNO) + (0,1) $) -- ($ (DP) + (-1,0) $);
			\draw[stealth-stealth, black!60!green] ($ (MNO) + (1,0) $) -- ($ (MNO) + (5,0) $) -- ($ (SR) + (0,-0.5) $);
			
			\draw[black!60!blue, <<->>, line width=0.3mm] ($ (MNO) + (0,-0.5) $) -- ($ (MNO) + (0,-1.5) $) -- node[above] {(*)} ($ (P1) + (0,-1) $) -- ($ (P1) + (0,-0.3) $);

			\draw[double, stealth-stealth, black!60!green, line width=0.7mm] (0,-2) -- (2,-2);
			\node[right] at (2.5,-2) {Main eUICC over the air channel}; 
			\draw[stealth-stealth, dashed] (0,-2.5) -- (2,-2.5);
			\node[right] at (2.5,-2.5) {eUICC intra communication using GlobalPlatform Open framework};			
			\draw[stealth-stealth, black!60!green] (0,-3) -- (2,-3);
			\node[right] at (2.5,-3) {MNO, SM-DP and SM-SR communications for eUICC management};				
			\draw[black!60!blue, <<->>, line width=0.3mm] (0,-3.5) -- (2,-3.5);
			\node[right] at (2.5,-3.5) {MNO access to \textbf{its} profile for connectivity parameters and Policy updates};
			\node at (1,-4) {\includegraphics[width=18pt]{certif.png}};
			\node[right] at (2.5,-4) {Entity certified by an authority};
			\node at (1,-4.5) {(*)};
			\node[right] at (2.5,-4.5) {Message respecting the Application Protocol Data Unit (APDU) format};		
			
  		\end{tikzpicture}
	}
	\caption{eUICC remote provisioning interfaces and communication channels}\label{fig:remoteprovisioning}
\end{figure*}

In addition to remote provisioning and re-programmability, the eUICC architecture allows several profiles (each being an application containing the subscription data of an MFF2 SIM), owned by different MNOs, to coexist on the same card during their lifetime.
This improvement eases the embedding of eUICCs inside devices during manufacture.
By comparison, MFF2 SIM cards are restricted to a single SIM application, owned by a single MNO.
Due to this distinction, the role of the MNO for eUICCs compared to its role for MFF2 SIMs is revised, in particular, the management of eUICCs is pushed from MNOs to the subscription manager described in the following paragraph.

\iftoggle{ArxivPaper}{
	\subsection{Subscription manager}
}
\paragraph{Subscription manager.}
The subscription manager's duties are split between two sub-entities: the \emph{\ac{dp}} and 
the \emph{\ac{sr}} (introduced in Section 2).
\begin{itemize}
	\item The \ac{dp} is responsible for generating profiles containing an operator's subscription data, and for transferring such profiles onto the eUICC.\footnote{GSMA suggests that an MNO can also assume the role of a DP or can own a DP~\cite{GSMABPR}.}
	\item \ac{sr}s are the main entity communicating with eUICCs. 
They remotely manage eUICCs using an eUICC's ISD-R interface.
All remote communication is secured through the secure channel that the SM-SR set up with the ISD-R. 
\end{itemize}

\iftoggle{ArxivPaper}{
	\subsection{SM-SR handover}
}

\paragraph{Secure Channels.}
To ensure confidentiality of messages exchanged, secure channels are used for communication between the eUICC and an external entity.
In order to achieve message confidentiality and maintain those secure channels, remote provisioning relies on protocols defined by ETSI \cite{ETSI102225,ETSI102226}, in particular, the \emph{Secure Channel Protocol} $80$ (SCP80).
This includes communication between the SM-DP and the ISD-P during profile installation (see Section \ref{sec:profile}), but also communication between an MNO and its profile (as represented in Figure \ref{fig:remoteprovisioning}).
However it is important to note that the mentioned secure channels are built upon the secure channel between the ISD-R and the SM-SR.

\iftoggle{ArxivPaper}{
	Another communication channel, relying on the SCP80 protocol, exists between an MNO and its profile on the eUICC.
	This channel allows the MNO to modify profile parameters, such as the NAA parameters.
	\begin{figure}[!ht]
		\centering
	    \resizebox{\textwidth}{!}{
			\begin{tikzpicture}[font=\fontsize{12}{22.4}\selectfont]
		  		\node (SR) at (0,0) {};
		  		\node (R) at (7,0) {};
		  		\node (DP) at (-5,0) {};
		  		\node (P) at (10.5,0) {};

		  		\draw [fill = black!30] ($ (R) + (-2,2.5) $) -- ($ (P) + (2.5,2.5) $) -- ($ (P) + (2.5,-2.4) $) -- ($ (R) + (-1.5,-2.4) $) -- ($ (R) + (-2,-1.9) $) -- cycle;
				\node[right, below] at ($ (R) + (-1,2.2) $)  {\Large{\textit{eUICC}}};	  		
	  		
		  		\draw (SR) -- ($ (SR) + (0,-2) $);
		  		\draw (R) -- ($ (R) + (0,-2) $);
		  		\draw (DP) -- ($ (DP) + (0,-2) $);
		  		\draw (P) -- ($ (P) + (0,-2) $);
		  		\draw[below right,font=\fontsize{16}{22.4}\selectfont] node at ($ (SR) + (0,-1.5) $) {- $k_{80}$};
		  		\draw[below right,font=\fontsize{16}{22.4}\selectfont] node at ($ (R) + (0,-1.5) $) {- $k_{80}$};
		  		\draw[below right,font=\fontsize{16}{22.4}\selectfont] node at ($ (P) + (0,-1.5) $) {- $k$};
		  		\draw[below right,font=\fontsize{16}{22.4}\selectfont] node at ($ (DP) + (0,-1.5) $) {- $k$};
				\draw (DP) node[abstract2] {SM-DP};
				\draw (SR) node[abstract2] {SM-SR};
				\draw (R) node[abstract2] {ISD-R};
				\draw (P) node[abstract2] {ISD-P};

				\node [cylinder,draw=black,thick,aspect=0.5,minimum height=5cm,minimum width=2cm,shape border      
			rotate=0,cylinder uses custom fill, cylinder body fill=red!30,cylinder end fill=red!5, opacity=0.8] at (3.5,0) {};

		   		\node [cylinder,draw=black,thick,aspect=0.5,minimum height=3cm,minimum width=2cm,shape border      
			rotate=0,cylinder uses custom fill, cylinder body fill=red!30,cylinder end fill=red!5, opacity=0.8] at (-2.5,0) {};

	   			\node [cylinder,draw=black,thick,aspect=0.5,minimum height=13.3cm,minimum width=0.5cm,shape border      
	   		rotate=0,cylinder uses custom fill, cylinder body fill=black!40!green,cylinder end fill=green!5, opacity=0.8] at (2.7,0.5) {};

				\draw[-stealth, black] ($ (DP) + (0.8,0.5) $) -- ($ (P) + (-0.6,0.5) $);
				\draw[red] (3.5,-0.5) node {SCP80};
				\draw[red] (-2.5,-0.5) node {Secure Channel};
		  	\end{tikzpicture}
		}
		\caption{Interface between the eUICC and the \ac{dp}. $k$ and $k_{80}$ are secret keys. The former one is obtained from a key agreement between the SM-DP and the eUICC 	(see Figure \ref{fig:keyexchange}), while the later one respects the requirements of SCP80.} \label{fig:es8}
	\end{figure}
}

\paragraph{SM-SR handover.}
Similarly to MFF2 SIMs that are managed by a single MNO, eUICCs are managed by a single \ac{sr}, albeit, the SM-SR may change over time.
EUM send an eUICC's data to the first SM-SR responsible for that eUICC.
This data includes the \texttt{EIS} file and the secret key needed to communicate with the eUICC.
\iftoggle{ArxivPaper}{
	This data includes the \texttt{EIS} file and the secret key $k_{80}$ needed to communicate with the eUICC.
}
After that, MNOs might request a change of the SM-SR responsible for an eUICC when a change of subscription occurs from one MNO to another.\\

\iftoggle{ArxivPaper}{
	Remote provisioning permits this change, and GSMA's specification presents the procedure used in such a case:
	\begin{enumerate}
		\item An MNO initiates \sout{the}a change request and notifies the new SM-SR of the incoming change.
		\item The new SM-SR verifies that it can support management of the eUICC.
		\item The MNO requests the current SM-SR to start the \texttt{SM-SR Change} procedure.
		\item The current SM-SR hands over the \texttt{EIS} file to the new SM-SR.
		\item This new SM-SR executes a key establishment procedure with the eUICC (similar to the one described in Section \ref{sec:download}) to obtain a shared key $k_{80}$ with the eUICC.
		\item Once the new secret key is present on the eUICC, the eUICC deletes the old SM-SR key.
		\item The old SM-SR deletes the EIS file of the eUICC.  
	This final step completes the transfer of ownership of an eUICC from one SM-SR to another.\\
	\end{enumerate}
}

\iftoggle{ArxivPaper}{
	\subsection{Secure channel protocol}
	Remote provisioning involves transmitting data over the air.
	Confidentiality of this data is paramount for MNOs whose secrets within profiles are uploaded inside eUICCs.
}
\label{sec:channelproto}

\noindent GSMA's specification describes the different communication channels and messages sent between the eUICC and its ecosystem, as depicted in Figure \ref{fig:remoteprovisioning}.
The specification also outlines the profile life cycle induced by the remote provisioning protocols, which we will describe next.
\section{Profile lifecycle}
\label{sec:profile}

Upon receipt of an eUICC equipped device, a customer can subscribe to an MNO.
To enable the contracted services, the MNO must upload a profile\footnote{In GSMA's specification, this profile is referred to as an \emph{operational profile}, to differentiate it from a provisioning profile.} corresponding to the subscription made onto eUICCs.
The upload and installation of a profile on an eUICC is as follows:

\begin{enumerate}
\item The MNO prepares a profile description containing the profile type and \emph{eUICC identifier} (EID), and sends the description to the \ac{dp} within a \texttt{DownloadProfile} request.
\item The SM-DP creates a profile corresponding to the request and containing the subscription data of the operator, then requests the creation of an ISD-P to the \ac{sr}.
\item The \ac{sr} instructs the ISD-R to create a new ISD-P capable of managing the profile created.
\item The \ac{dp} derives a secret key $k$ from a shared secret $s$ with the \ac{isdp}.
The secret $s$ is obtained by following the Elliptic Curve Key Agreement (ECKA) protocol based on ElGamal \cite{GPCS}. 
\iftoggle{ArxivPaper}{
	GSMA uses the Elliptic Curve Key Agreement (ECKA) protocol based on ElGamal \cite{GPCS} which is represented by Figure \ref{fig:keyexchange} and described here:
	\footnote{To ease the comprehension of the paper, the protocol described here omits details of an eUICC's internal mechanisms, and of the key derivation protocol used.}
	\textcolor{blue}{I don't see a reason to describe this in detail. We don't describe other cryptographic operations in such detail.}
	\begin{enumerate}
		\item The \ac{dp} gets the certified public key of the eUICC from the \texttt{EIS} file present in the \ac{sr},
		\item The \ac{dp} sends its certified public key to the eUICC.
		Using the certificate authority public key present in the ECASD, the ISD-P is able to verify this public key.
		\item The eUICC responds with a random challenge to the \ac{dp},
		\item The \ac{dp} creates an ephemeral public key pair and sends the public part to the eUICC, along with the challenge previously received and the ephemeral key pair signed using its private key,
		\item Both sides obtain a shared secret $s$ by combining both the ephemeral key pair and the eUICC key pair; and then, they derive a key $k$ (called the \emph{Profile Management Credentials}) from this shared secret.
	\end{enumerate}
}
\item After that, the profile, created by the SM-DP, is sent by the SM-DP to the eUICC over a secure channel set up using $k$ (see Section \ref{sec:remoteProvisioning}\iftoggle{ArxivPaper}{ and illustrated in Figure \ref{fig:es8}}).
\item Finally, the profile is uploaded by the SM-DP onto the eUICC.
\item The ISD-P decrypts the profile using $k$ and installs the profile.
\end{enumerate}

%
%
%
%

Installed profiles are initially disabled and may subsequently be enabled or deleted.
To manage profiles, GSMA's specification defines a set of \emph{policy rules}, \texttt{POL1}, that are stored in a file inside each profile.
Policy rules are instantiated by MNOs during profile creation and can be updated remotely by the MNO for the enabled profile only.
For this purpose, MNOs can set up a secure channel with the profile using a security domain, the \emph{MNO-SD}, present in the profile.
To ease profile management, an unsynchronized version of those rules are maintained by the MNO in the EIS file (held by the SM-SR).
Policy rules determine if a profile can be disabled, deleted or must be deleted once it is disabled.
To ensure exclusivity, operators can lock devices to their profile.
Indeed policy rules can prevent the change of a subscription by locking the device to the enabled profile -- similarly to the SIM lock feature of mobile phones.
M2M devices can only access the network with the profile that is enabled.

GSMA, in order to increase the reliability of eUICCs connectivity, also introduces another attribute for the profiles: the \emph{fallback attribute}.
This Boolean attribute can only be set to true on a single disabled profile which is enabled to prevent connection loss or errors. 

GSMA choses to rely on profiles to improve over MFF2 SIMs, their lifecycle enabling re-programmability support remote provisioning. Furthermore the format of a profile eases the isolation of sensitive data ensuring the security requirements of the different stakeholders.





\section{Closing remarks}

The MFF2 form factor requires M2M SIMs be embedded inside devices.
This presents some shortcomings, especially as such SIMs support only a single subscription that must be physically uploaded. New technologies have been explored to overcome these shortcomings. In particular, 
GSMA has released a specification describing mechanisms for remote provisioning, and their application to the management of a new SIM, called an eUICC, that is re-programmable.
eUICCs can be embedded in M2M devices 
during manufacture, 
and need not be replaced upon a consumer changing subscriptions. 


This article presents the motivation behind eUICC and remote provisioning, and the changes they will induce upon the telecommunication ecosystem.
It also describes 
the technical challenges of remote provisioning and illustrates GSMA's 
solution.

eUICC is being driven towards standardization. 
If eUICCs are adopted widely as the next generation SIMs for M2M devices, then, operators, device manufacturers and SIM manufacturers will have to adapt to the new eUICCs ecosystem.
The evolution towards next generation telecommunications is exciting, but not without risk; flawed systems have the potential to cost society dearly, in terms of both lost liberties and financial costs.
As such, similarly to the security analysis done by Meyer et al. in \cite{meyer2018attacks}, it will be important to analyze in detail the mechanisms behind remote provisioning to be sure that the properties offered by the current model will hold for eUICCs, especially in term of security as it is paramount for operators. 
Such an overview is helping in that regard by easing the comprehension of those technical specification.
Furthermore eUICCs are also likely to be considered as the next generation SIM for \emph{all} devices, in particular Internet of Things devices considered as potential M2M devices \cite{iot_esim_gsma} and mobile phones already marketed as an eSIM by GSMA \cite{esim_gsma}.
This represents a very important market, this is why it is crucial to look at the impact that eUICCs and remote provisioning can have, both in terms of changes to the network and to consumer habits.
	
\paragraph{Acknowledgements.} We are grateful to Aisha Chudasama Mahmud for feedback that helped improve this article.



\bibliographystyle{splncs03}
\bibliography{bibfile}

\end{document}